\documentclass[
    ,final            
  ]
  {aipproc}

\layoutstyle{6x9}

\begin{document}

\title{Effects of quark matter and color superconductivity in compact stars}

\author{D. Blaschke}{
  address={Fachbereich Physik, Universit\"at Rostock, D-18051 Rostock,
Germany}
	,altaddress={Bogoliubov  Laboratory of Theoretical Physics,\\
        Joint Institute for Nuclear Research, 141980, Dubna, Russia}
}

\author{H. Grigorian}{
  address={Fachbereich Physik, Universit\"at Rostock, D-18051 Rostock,
Germany}
	,altaddress={Department of Physics, Yerevan State University,
375025 Yerevan, Armenia}
}	
\author{D.N. Aguilera}{
  address={Fachbereich Physik, Universit\"at Rostock, D-18051 Rostock,
Germany}
  ,altaddress={Instituto de F\'{\i}sica Rosario, Bv. 27 de febrero 210 bis,
        2000 Rosario, Argentina} 
}
\author{S. Yasui}{
  address={Fachbereich Physik, Universit\"at Rostock, D-18051 Rostock,
Germany}
  ,altaddress={Research Center for Nuclear Physics, Osaka University, 
Ibaraki 567 - 0047, Japan} 
}
\author{H. Toki}{
  address={Research Center for Nuclear Physics, Osaka University, 
Ibaraki 567 - 0047, Japan}
}

\begin{abstract}
The equation of state for quark matter is derived for a nonlocal, chiral quark
model within the mean field approximation.
We investigate the effects of a variation of the form factors of the 
interaction on the phase diagram of quark matter under the condition of 
$\beta$- equilibrium and charge neutrality.
Special emphasis is on the occurrence of a diquark condensate which signals a
phase transition to color superconductivity and its effects on the equation of
state.
We calculate the quark star configurations by solving the Tolman- Oppenheimer-
Volkoff equations and obtain for  
the transition from a hot, normal quark matter core of a protoneutron star 
to a cool diquark condensed one a release of binding energy of the order of 
$\Delta M c^2 \sim 10^{53}$ erg.
We study the consequences  of antineutrino trapping in hot quark matter
for quark star configurations with possible diquark condensation and discuss 
the claim that this energy could serve as an engine for explosive
phenomena.
A "phase diagram" for rotating compact stars (angular velocity-baryon mass
plane) is suggested as a heuristic tool for obtaining constraints on the
equation of state of QCD at high densities. It has a
critical line dividing hadronic from quark core stars which is
correlated with a local maximum of the moment of inertia and can thus be
subject to experimental verification by observation of the rotational
behavior of accreting compact stars.
\end{abstract}

\maketitle


\section{Introduction}
Color superconductivity in quark matter \cite{Rajagopal:2000wf} is one 
interesting aspect of
recent discussions devoted to the physics of compact star interiors
 \cite{Blaschke:2001uj}.
Since calculations of the energy gap of quark pairing predict
a value  $\Delta \sim 100$ MeV and corresponding critical
temperatures for the phase transition to the superconducting state are 
expected to follow the BCS relation $T_c = 0.57 ~\Delta$, the question arises 
whether diquark condensation can lead to remarkable 
effects on the structure and evolution of compact objects.
If positively answered, color superconductivity of quark matter could 
provide signatures for the detection of a deconfined phase in the
interior of compact objects (pulsars, Low-mass X-ray binaries)
 via observations.
Hong, Hsu and Sannino have conjectured \cite{Hong:2001gt} 
that the release of binding energy due to Cooper pairing of quarks in 
the course of protoneutron star evolution could provide an explanation
for the unknown source of energy in supernovae, hypernovae or gamma-ray 
bursts, see also \cite{Ouyed:2001bm}. 
Their estimate of energy release did not take into account the change in 
the gravitational binding 
energy due to the change in the structure of the stars quark core.
We reinvestigate the question of a possible binding 
energy release due to a color supercondcutivity transition by taking into 
account changes in the equation of state (EoS) and the configuration of the
quark star selfconsistently and by including the effects of antineutrino
trapping \cite{Aguilera:2002dh}.

As a first step in this direction we will discuss here
the two flavor color superconducting (2SC) quark matter
phase which occurs at lower baryon densities than the
color-flavor-locking (CFL) one, see \cite{Steiner:2002gx,Neumann:2002jm}.
We will investigate the influence of the formfactor of the interaction 
on the phase diagram and the EoS of dense quark matter
under the conditions of charge neutrality and
isospin asymmetry due to $\beta$-equilibrium relevant for compact stars.

Finally we consider  the question whether the effect of
diquark condensation which occurs in the earlier stages of the compact star
evolution ($t \simeq 100$ s) \cite{Blaschke:2000dy,Carter:2000xf} at 
temperatures $T \sim T_c \sim 20\div 50$ MeV
can be considered as an engine for exposive astrophysical phenomena like 
supernova explosions due to the release of a binding energy of
about $10^{52} \div 10^{53}$ erg, as has been suggested before 
\cite{Hong:2001gt,Ouyed:2001bm}.

\section{Thermodynamics of a nonlocal chiral quark model}
The phase structure of electrically and color neutral quark matter 
in $\beta$-equilibrium 
has been 
studied in \cite{Neumann:2002jm} and 
it has been shown that at densities relevant for compact star interiors
the 2SC phase should be dominant over the CFL phase. 
The latter one could only be stable in the very inner core and 
thus  does not occupy a large enough volume in 
order to cause observable effects.
Therefore, we consider in the present work two flavor quark matter in 
with 2SC superconductivity only.
The more general case which includes the CFL phase does not invalidate the 
scenario developed in the following and will be studied in a subsequent work. 
We consider the grand canonical thermodynamical potential 
for 2SC quark matter within a nonlocal chiral quark model 
\cite{Blaschke:2003yn} 
where in the mean field approximation the mass gap $\phi$ and the 
diquark gap $\Delta$ appear as order parameters which can be expressed as in 
\cite{Kiriyama:2001ud} by
\begin{eqnarray} \label{Omeg1}
\Omega_q(\phi,\Delta;\mu_q,\mu_I,T)&=&
\frac{\phi^2}{4G_1}+\frac{\Delta^2}{4G_2}
-\frac{1}{\pi^2}\int^\infty_0dqq^2\{
\omega\left[\epsilon_r(-\mu_q-\mu_I),T\right]
\nonumber\\ 
& &+
\omega\left[\epsilon_r(\mu_q-\mu_I),T\right]+
\omega\left[\epsilon_r(-\mu_q+\mu_I),T\right]+
\nonumber\\ 
& &+
\omega\left[\epsilon_r(\mu_q+\mu_I),T\right]
\}
-\frac{2}{\pi^2}\int^\infty_0dqq^2\{
\omega\left[\epsilon_b(E(q)-\mu_q)-\mu_I,T\right]
\nonumber\\
& &
+ \omega\left[\epsilon_b(E(q)+\mu_q)-\mu_I,T\right]+
\omega\left[\epsilon_b(E(q)-\mu_q)+\mu_I,T\right]
\nonumber\\
& &
+\omega\left[\epsilon_b(E(q)+\mu_q)+\mu_I,T\right]
\}
+\Omega_{vac}~,
\label{ome9}
\end{eqnarray}
where we have introduced   
the quark chemical potential $\mu_q=(\mu_u+\mu_d)/2$ and the 
chemical potential of the isospin asymmetry $\mu_I = (\mu_u-\mu_d)/2$
instead of the chemical potentials of up and down quark flavors.
We neglect here a possible difference between the chemical potentials of 
paired and unpaired colors for the same quark flavor. For a more general
approach see \cite{Neumann:2002jm,Huang:2002zd}.
The factor $2$ in the last integral comes from the degeneracy of the
blue and green colors ($\epsilon_b=\epsilon_g$). 
We have introduced the notation  
\begin{eqnarray}
&&
\omega\left[\epsilon_c,T\right]= T\ln\left
[1+\exp\left(-\frac{\epsilon_c}{T}\right)\right]+\frac{\epsilon_c}{2}~,
\label{ome3}
\end{eqnarray} 
where the first argument is  given by
\begin{eqnarray}
&&\epsilon_c(x)=x\sqrt{1+\Delta^2_c/x^2}~,
\end{eqnarray}
with the color index $c=r,g,b$ and we assume that 
the green and blue colors  are paired
while the red one remains unpaired, so that we have 
\begin{eqnarray}
&& \Delta_c=g(q)\Delta(\delta_{c,b}+\delta_{c,g}).
\end{eqnarray}
The dispersion relation for unpaired quarks with dynamical mass
function $m(q)=m+g(q)\phi$
is given by
\begin{eqnarray}
&& E_f(q)=
\sqrt{q^2+m^2(q)}~,
\end{eqnarray} 
where $g(q)$ denotes the formfactor of the quark interaction, for which we 
employ the following models
\begin{eqnarray}
\label{LF}
g_{\rm L}(q) &=& [1 + (q/\Lambda_{\rm L})^{2\alpha}]^{-1},~~~~ \alpha > 1~,
\\
\label{GF}
g_{\rm G}(q) &=& \exp(-q^2/\Lambda_{\rm G}^2)~,
\\
\label{NF}
g_{\rm NJL}(q) &=& \theta(1 - q/\Lambda_{\rm NJL})~.
\end{eqnarray}
Depending on the parameter $\alpha$, the Lorentzian (L) momentum distribution 
can interpolate between a soft Gaussian (G) formfactor (for $\alpha \sim 2$) 
and a hard cutoff (NJL) one (for $\alpha > 30$).
The parametrization of the model can be found in Refs. 
\cite{Schmidt:1994di,Blaschke:2003yn}.
The contribution from the leptons should be added to the quark thermodynamical 
potential $\Omega_q$ in order to obtain the total one
\begin{eqnarray}
\Omega(\phi,\Delta;\mu_q,\mu_I,\mu_e,\mu_{\bar \nu_e},T)&=& 
\Omega_q(\phi,\Delta;\mu_q,\mu_I,T) 
+ \sum_{l \in \{e,\bar \nu_e \}}\Omega^{id}(\mu_l,T)~,
\end{eqnarray}
where
\begin{eqnarray}
&&
\Omega^{id}(\mu,T)=-\frac{1}{12\pi^2}\mu^4-\frac{1}{6}\mu^2T^2-\frac{7}{180}
\pi^2T^4
\end{eqnarray}
is the thermodynamical potential for an ideal gas of massless 
fermions.
The stellar matter in the quark core of compact stars consists of 
$u$ and  $d$ quarks, electrons $e$ and antineutrinos  $\bar \nu_e$ 
under the conditions of
$\beta$-equilibrium:  
$d \longleftrightarrow u+e^-+\bar \nu_e$,
which in terms of chemical potentials reads
$
\mu_e+\mu_{\bar \nu_e}=-2\mu_I~,
$
and charge neutrality: $\frac{2}{3}n_u-\frac{1}{3}n_d-n_e = 0$,
which also could be written as
$
n_q+3n_I-6n_e=0~.
$
The number densities $n_j$, $j\in \{u,d,e,\bar \nu_e,q,I\}$ are defined as
\begin{eqnarray}
&&
n_j=-\frac{\partial \Omega}{\partial \mu_j}\bigg|_{\phi_0,\Delta_0;T}.
\end{eqnarray}
The conditions for the local extrema of $\Omega_q$, correspond to 
coupled gap equations for the two order parameters $\phi$ and  $\Delta$
\begin{eqnarray}
&&
{\partial \Omega \over \partial \phi}\bigg|_{\phi=\phi_0,\Delta=\Delta_0}=
{\partial \Omega \over \partial \Delta}\bigg|_{\phi=\phi_0,\Delta=\Delta_0}=0~.
\label{gaps}
\end{eqnarray}
The global minumum of $\Omega_q$ represents the state of thermodynamical 
equilibrium from which all equations of state can be obtained by derivation.
In the following subsections we want to comment on aspects of this
stellar matter model which turns out to be essential for the discussion of 
quark matter in compact stars.

\subsection{Effect of formfactors}
\begin{figure}[htb]
  {\includegraphics[width=.5\textwidth,angle=-90]{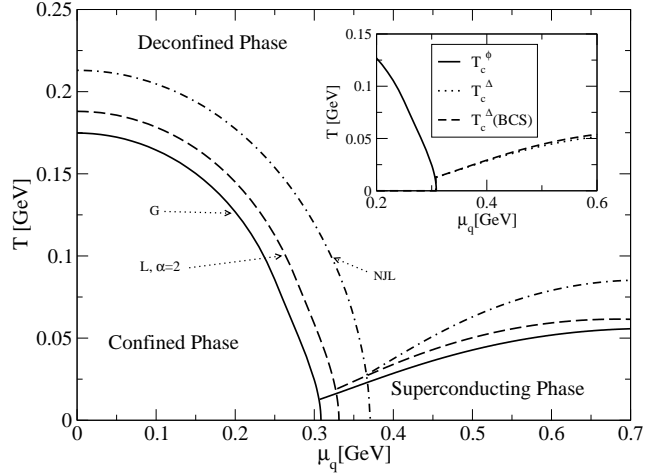}}
  \caption{Phase diagrams for different
form factors: Gaussian (solid lines), Lorentzian $\alpha=2$ (dashed
lines)  and NJL
(dash-dotted). In the upper panel the comparison with the BCS formula for
$T_c = 0.57~ \Delta(T=0,\mu_q)~g(\mu_q)$ is shown for the Gaussian model.}
\label{PD}
\end{figure}

The nonvanishing of the order parameters $\phi$ or $\Delta$ signals the 
presence of a phase with broken chiral symmetry or color superconductivity,
respectively. In Fig. \ref{PD} we show the resulting phase diagram of 
quark star matter under the above constraints and neglecting the CFL phase
which should appear only at such high densities that it will at best occupy a 
negligible volume in the very inner core of a compact star configuration.  
From Fig. \ref{PD} we see that the softer the formfactor $g(q)$ the lower the 
critical temperatures and chemical potentials for the phase transition to 
quark matter with vanishing order parameters or to color superconducting 
quark matter at low temperatures. It is remarkable that a modified BCS relation
for the critical 2SC temperature holds \cite{Blaschke:2003yn}.  

\subsection{Antineutrino trapping}

\begin{figure}[htb]
\includegraphics[width=.5\textwidth,height=0.55\textwidth,angle=-90]{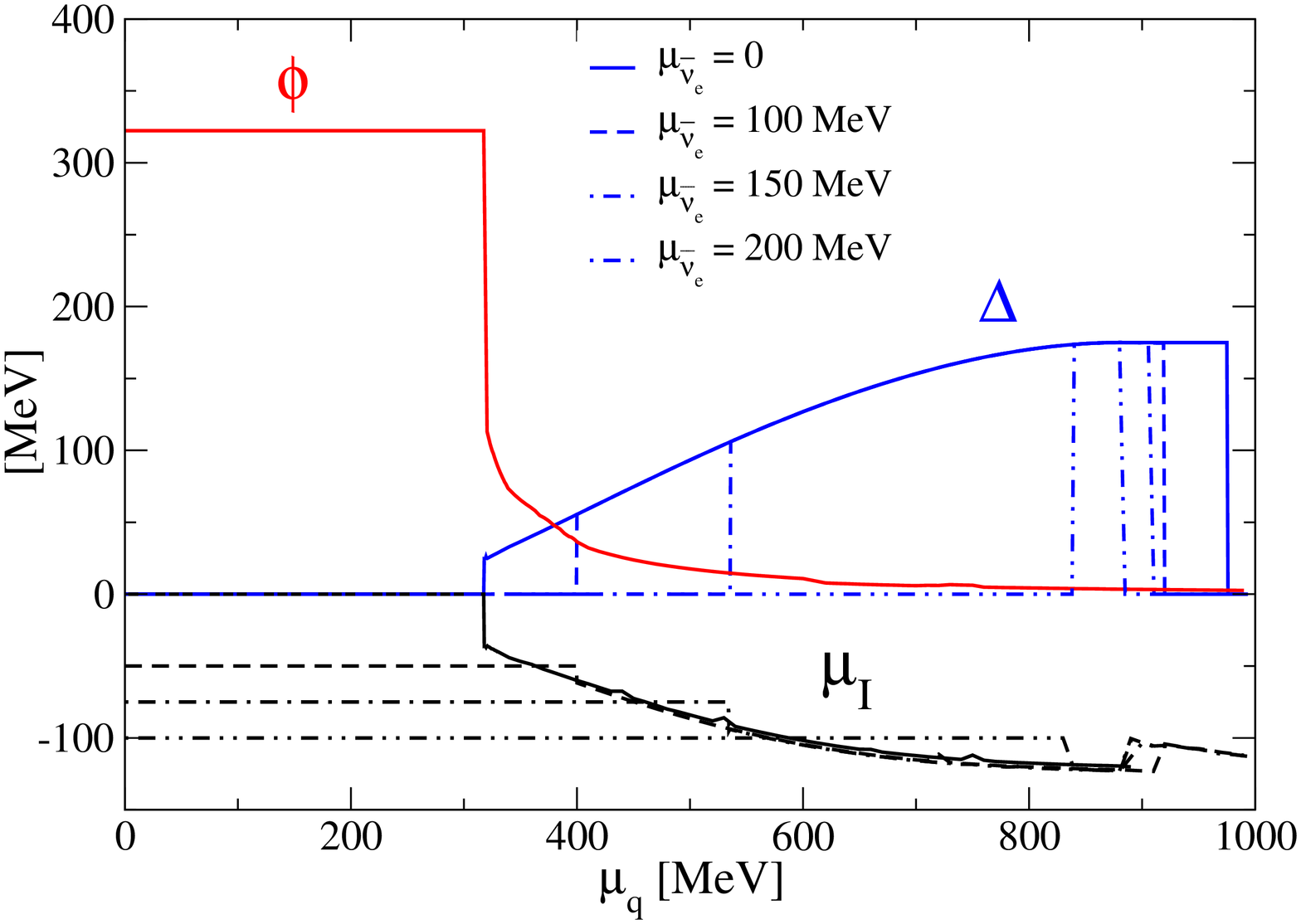}
\includegraphics[width=.5\textwidth,height=0.55\textwidth,angle=-90]
{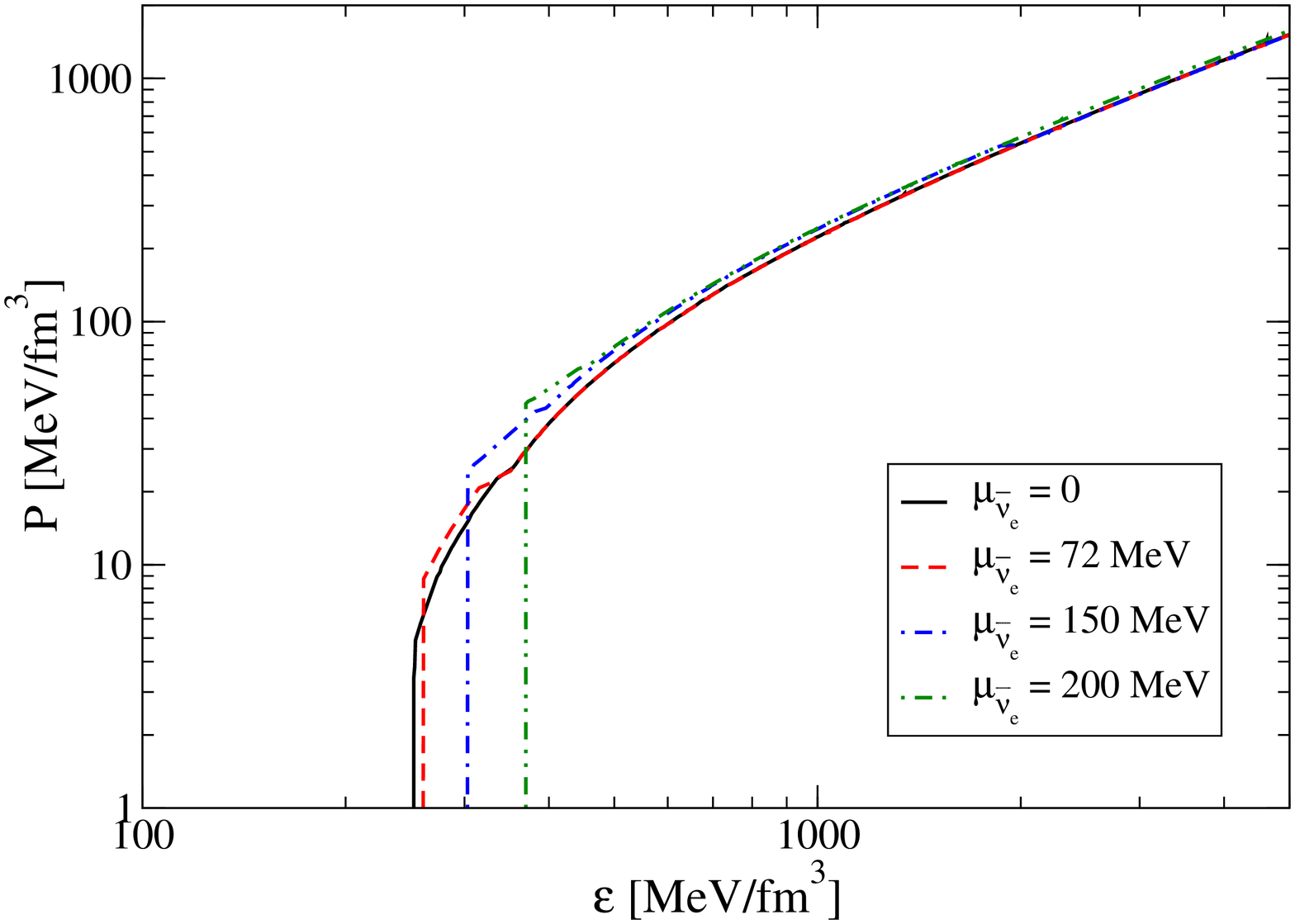}
\caption{Left panel: 
Mass gap $\phi$, diquark gap $\Delta$ and isospin chemical 
potential $\mu_I$ 
as a function of the quark chemical potential $\mu_q$ for different 
values of the antineutrino chemical 
potential $\mu_{\bar \nu_e}$. Solutions obey $\beta$-equilibrium and 
charge neutrality conditions.
Right panel: 
Pressure vs. energy density for different values of the antineutrino chemical 
potential $\mu_{\bar \nu_e}$. The onset of the appeareance of quark matter 
is shifted to higher energy densities due to  antineutrino trapping.}
\label{trap}
\end{figure}

The results for the solution of the gap equations (\ref{gaps}) are shown in
Fig. \ref{trap} (left panel) for different values of the antineutrino chemical 
potential
which is a measure for the density of trapped antineutrinos in a hot, young
protoneutron star. For values $\mu_{\bar\nu_e} > 72$ MeV  the flavor asymmetry 
becomes large enough to prevent diquark pairing and 
therefore color superconductivity at low densities.
Simultaneously, the onset density for quark matter occurence is shifted to
higher densities, see Fig. \ref{trap} (right panel). These solutions will be 
used for the discussion of a scenario of hot quark star evolution in the next 
Section.
Before that we have to consider the question whether the quark-hadron phase 
transition would occur at too high energy densities so that no quark core could
exist in a hybrid compact star.    

\subsection{Quark hadron phase transition}
We construct a quark hadron phase transition at zero temperature using 
a linear and a nonlinear Walecka model for the hadronic phase and perform a 
Maxwell construction for the phase transition.
Although it has been claimed that this procedure might be in contradiction with
the Gibbs conditions for phase equilibrium when more than one conserved charge 
exists in the system \cite{Glendenning:1992vb}, a recent investigation 
including charge screening has revealed the opposite 
\cite{Voskresensky:2001jq}.  
\begin{figure}[ht]
  {\includegraphics[width=.6\textwidth,angle=-90]{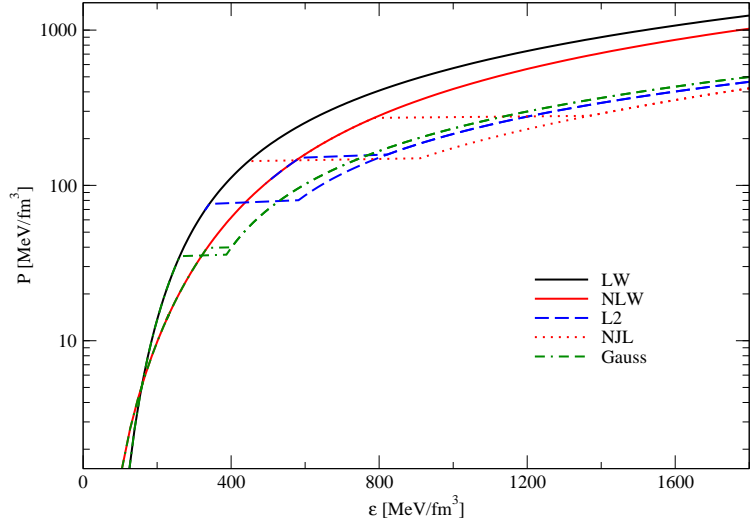}}
  \caption{EoS of compact star matter with quark hadron phase transition.
Hadronic EoS: linear (LW) and nonlinear (NLW) Walecka model, quark matter:
nonlocal separable model with 2SC quark matter and different form factors: 
Gaussian (Gauss), Lorentzian $\alpha=2$ (L2)  and cutoff (NJL). }
\label{EoSs}
\end{figure}
The resulting EoS with a deconfinement transition  
confirms that the lowest critical phase transition (energy) density is
obtained for the softest (Gaussian) formfactor model, see Fig. \ref{EoSs}.

\section{Compact star configurations}
What compact star configurations will result from the use of the above
EoS? Will a stable compact star with quark matter core be obtained?  
We will answer these questions in the present Section by solving the 
Tolman-Oppenheimer-Volkoff equations for the case without rotation and by
applying the perturbative method of solution of Einstein equations for 
stationary rigid rotation. First we consider 
quark stars without hadronic shell which can be thought of as the simplified 
models for a compact star interior and after that we discuss hybrid stars.  
\subsection{Quark stars - engine for explosive phenomena?}
The engine which drives supernova explosions and gamma ray bursts being
among the most energetic phenomena in the universe remains still puzzling.
The phase transiton to a quark matter phase 
may be a mechanism that could release such an amount of energy 
\cite{Drago:1997tn,Berezhiani:2002ks}.
It has been proposed that due to the Cooper instability in dense Fermi gas
cold dense quark matter shall be in the color superconducting state with
a nonvanishing diquark condensate \cite{Alford:2000sx,Blaschke:2001uj}. 
The consequences of diquark condensation for the cooling of compact stars
due to changes in the transport properties and neutrino emissivities 
have been investigated much in detail, see 
\cite{Blaschke:1999qx,Page:2000wt,Blaschke:2000dy,Blaschke:2003yn}, and may
even contribute to the explanation 
of the relative low temperature of 
the pulsar in the supernova remmant 3C58 \cite{cooling}.    
Unlike the case of normal
(electronic) superconductors, the pairing energy gap in quark matter is 
of the order of the Fermi energy so that diquark condensation gives
considerable contributions to the equation of state (EoS) of the order
of $({\Delta}/{\mu})^2$. Therefore, it has been suggested 
that there might be scenarios which 
identify the unknown source of the energy of $~10^{53}$ erg with a release 
of binding energy due to Cooper pairing of quarks in the core of a cooling 
protoneutron star \cite{Hong:2001gt}.    
In that work the total diquark condensation energy released in 
a bounce of the core is estimated as 
$({\Delta}/{\mu})^2M_{{\rm core}}$ corresponding to a few 
percent of a solar mass, that is $~10^{52}$ erg. It has
been shown in \cite{Blaschke:2003yn} 
by solving the selfconsistent problem of the star configurations, 
however, that these effects due to the softening of the EoS in the diquark
condensation transition lead to an increase in the gravitational mass 
of the star contrary to the naive estimates and that no explosion occurs.

\begin{figure}[ht]
\includegraphics[width=.45\textwidth,angle=-90]{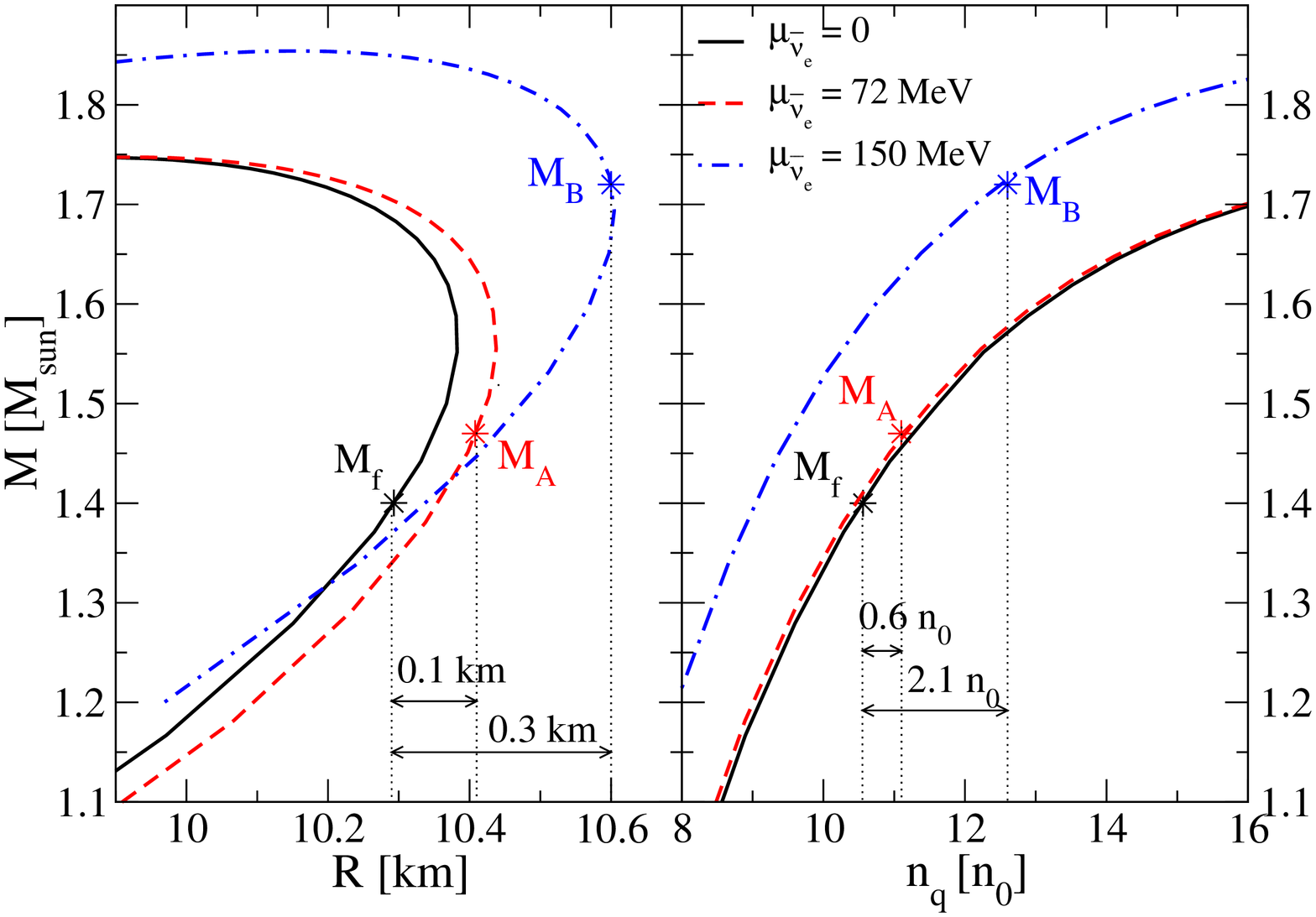}
\includegraphics[width=.45\textwidth,angle=-90]{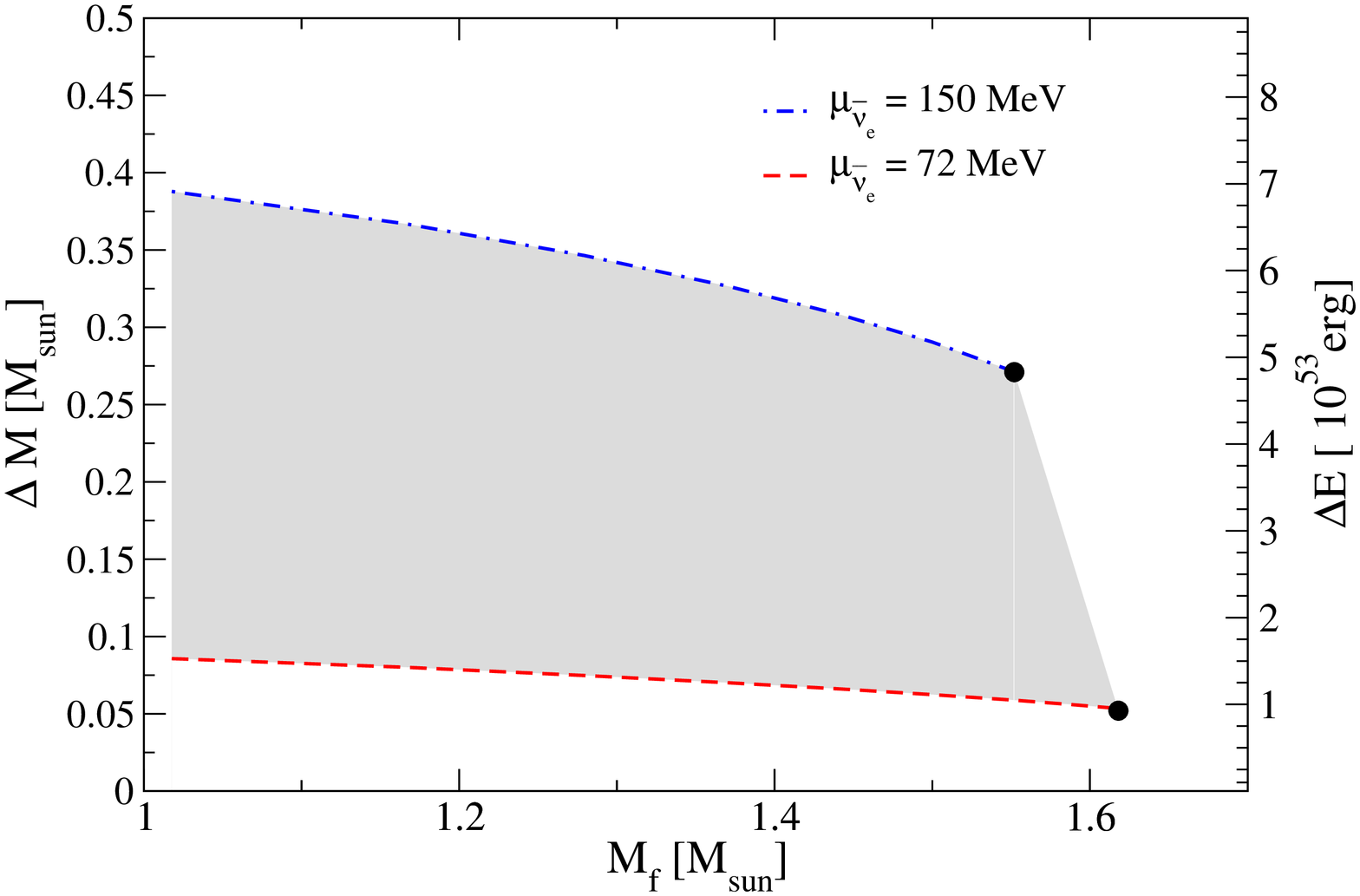}
  \caption{Left panel: Quark star configurations for different antineutrino 
chemical potentials $\mu_{\bar \nu_e}=0, ~72,~150$ MeV. 
The total mass $M$ in solar masses 
$M_{\odot}$ is shown as a function of the radius $R$ (left panel) and 
as a function of the central number density  $n_q$ in units of the 
nuclear number density $n_0$ (right panel). 
Asterisks denote  configurations with  the same total baryon number.
Right panel: Mass defect $\Delta M$ and corresponding energy release $\Delta E$
due to antineutrino untrapping 
as a function of the mass of the final state $M_f$.
The shaded region is defined by the estimates for the upper and lower
limits of the antineutrino chemical potential
in the initial state $\mu_{\bar \nu_e}=150$ MeV (dashed-dotted line) and 
$\mu_{\bar \nu_e}=72$ MeV (dashed line), respectively.}
\label{QSConf}
\end{figure}

Therefore, we have suggested a new  mechanism of energy release 
\cite{Aguilera:2002dh} which involves a first order phase transition induced 
by antineutrino untrapping. 
The trapping of antineutrinos occurs in hot compact star configurations at 
temperatures 
$T\geq 1$ MeV where the mean free path of (anti-)neutrinos becomes smaller
than the typical size of a star \cite{Prakash:2001rx}.  
\begin{figure}[h]
\includegraphics[width=.3\textwidth,angle=0]{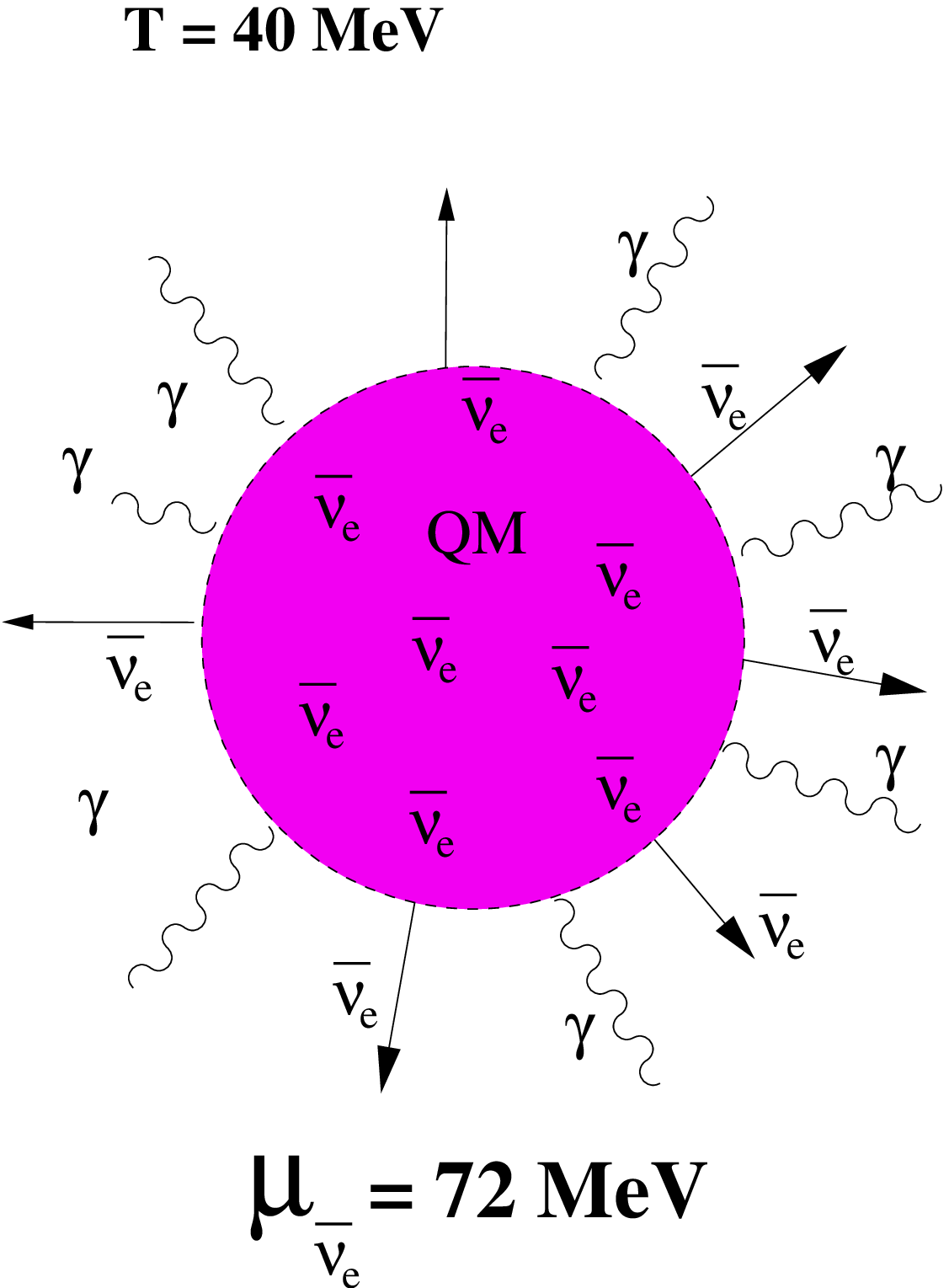}
\includegraphics[width=.3\textwidth,angle=0]{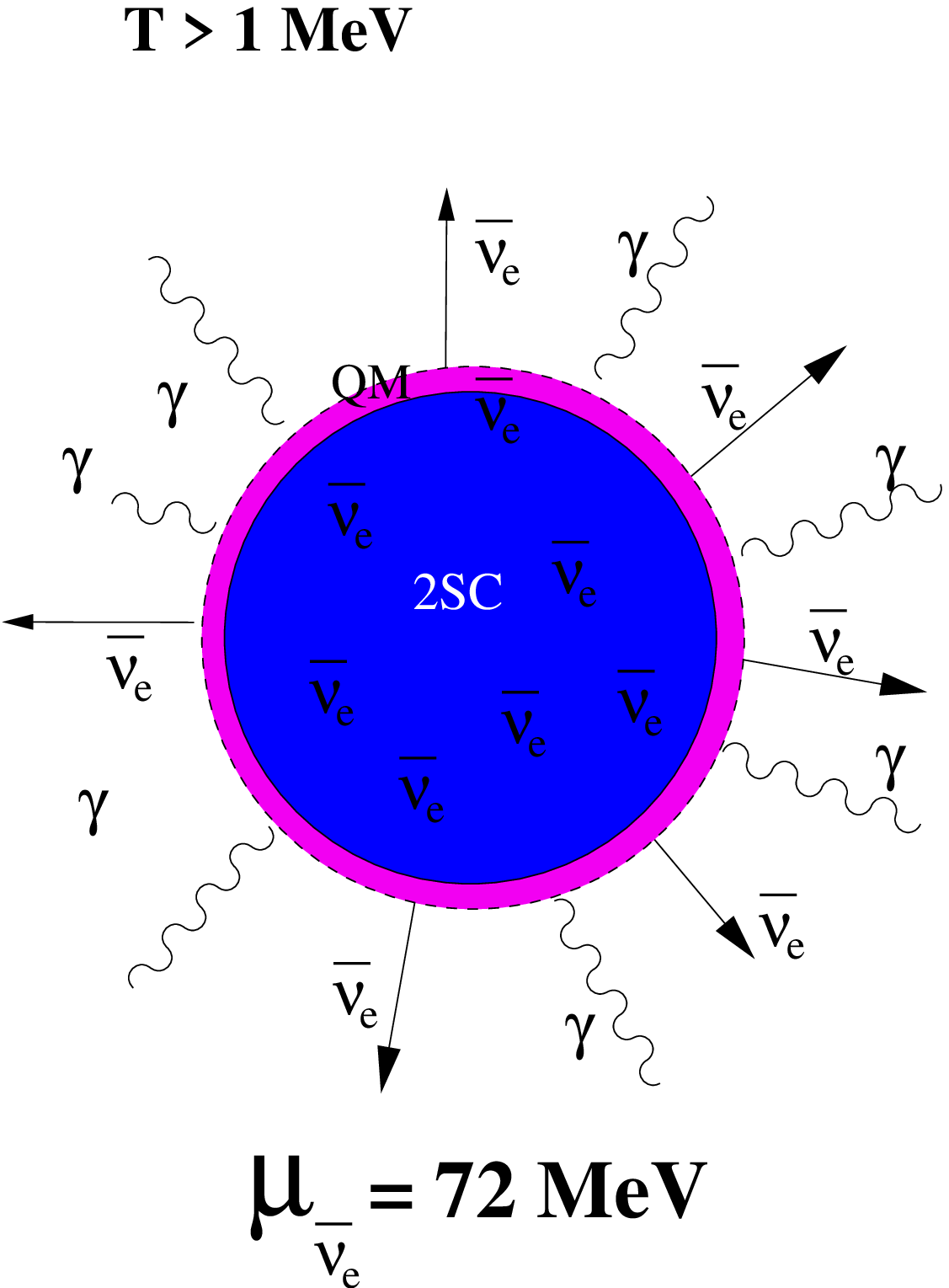}
\includegraphics[width=.3\textwidth,angle=0]{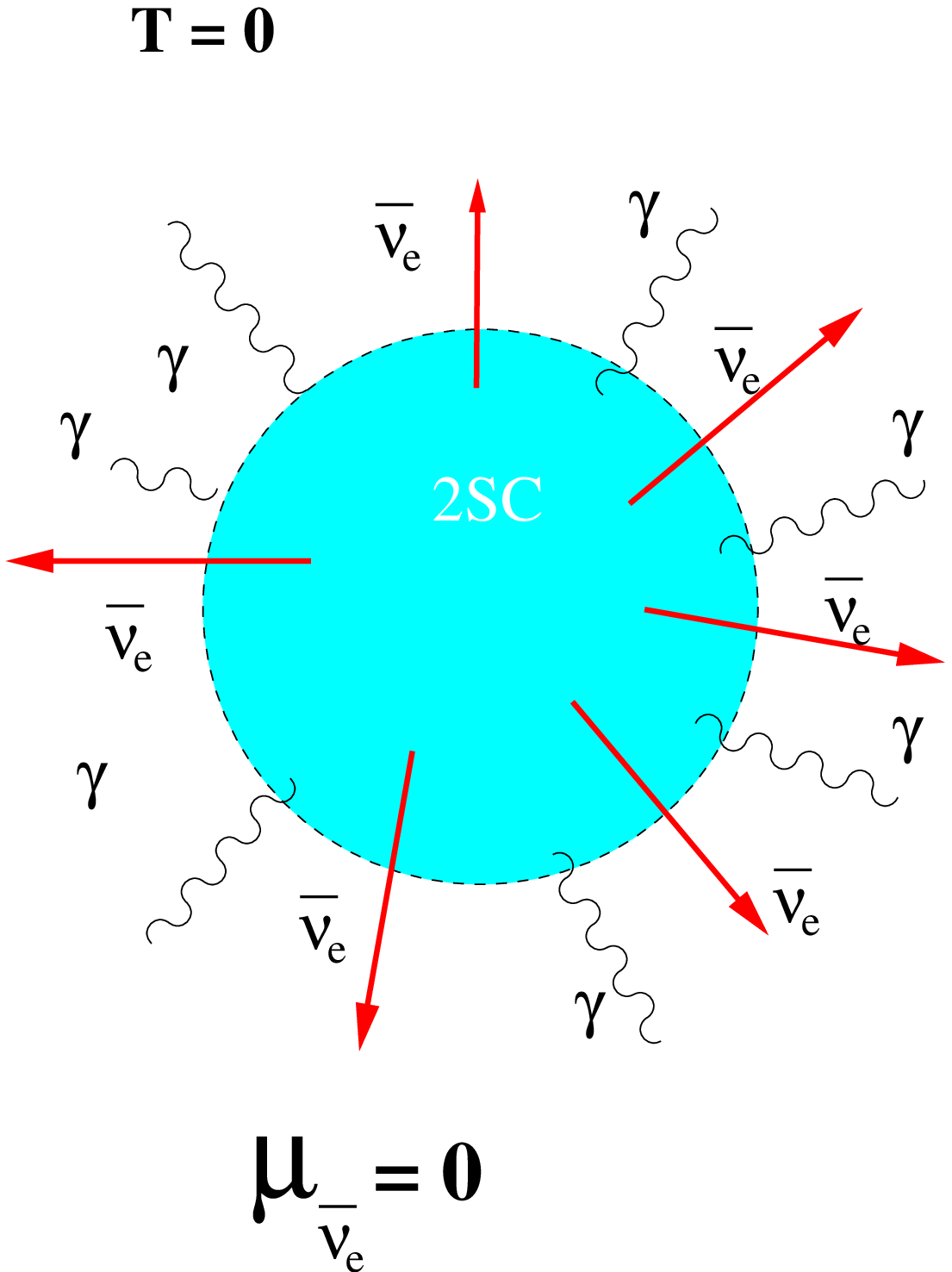}
\caption{Left two graphs: Quark star cooling by neutrino and photon 
emission from the surface in the case of antineutrino trapping when $T>1$ MeV.
Right graph: Antineutrino untrapping and burst-type release.}
\label{cooltrap}
\end{figure}
During the collapse in the  hot era of protoneutron star evolution, 
antineutrinos are produced due to the $\beta$-processes.
Since they have a  small mean free path, they cannot escape  
and the asymmetry in the system is increased.   
This entails that the formation of the diquark condensate
is shifted to higher densities or even inhibited depending on the 
fraction of trapped antineutrinos.
As the quark star cools, a two-phase structure will occur. 
Despite of the asymmetry, the interior of the quark star (because of its
large density) could consist of color superconducting quark matter, whereas
in the more dilute outer shell, diquark condensation cannot occur and quark 
matter is in the normal state, opaque to antineutrinos for $T\geq 1$ MeV.
When in the continued cooling process the antineutrino mean free path 
increases above the size of this normal
matter shell, an outburst of antineutrinos 
occurs and gives rise to an energy release of 
the order of $10^{53}-10^{54}$ erg. 
This untrapping transition is of first order and 
could lead to an explosive phenomenon.
Three stages of this scenario of hot quark star evolution are illustrated in 
Fig. \ref{cooltrap}. 
\begin{figure}[ht]
  {\includegraphics[width=.6\textwidth,angle=-90]{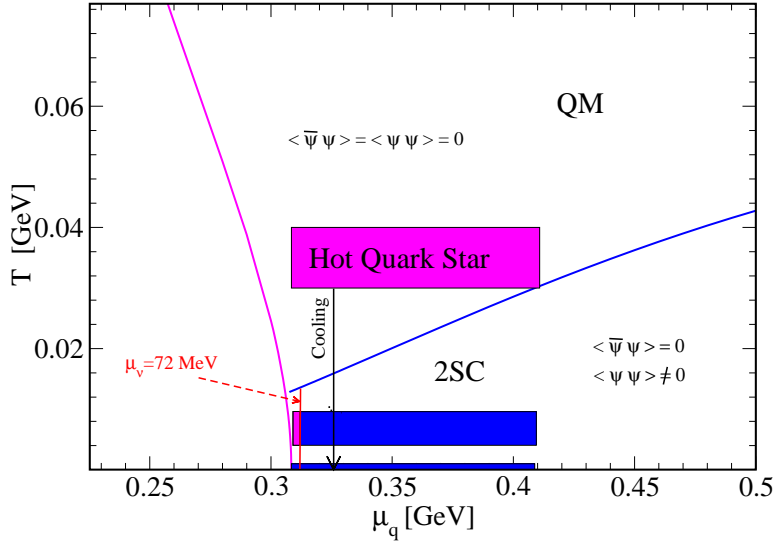}}
  \caption{Three stages of quark star cooling in the phase diagram 
corresponding to Fig. \ref{cooltrap}.}
\label{PDNT}
\end{figure}
In Fig. \ref{PDNT} these stages are shown schematically in the quark matter 
phase diagram. 
From Fig. \ref{trap} (right panel) we can see that the EoS without 
antineutrinos is softer than with antineutrinos (it has a lower pressure at a 
given energy density) and therefore allows more compact configurations 
(Fig. \ref{QSConf}, left part of left panel). 
The presence of antineutrinos tends to increase  the mass of the star for 
a given central density (Fig. \ref{QSConf}, right part of left panel).

To estimate the effect of antineutrinos on
star configurations we choose a reference configuration  without 
antineutrinos with the mass of a typical neutron star $M_f = 1.4~ M_{\odot}$ 
(see Fig. \ref{QSConf}). The corresponding radius is
$R_f = 10.29$ km and the central density $n_q = 10.5~n_0$, 
where $n_0 = 0.16~{\rm fm}^{-3}$
is the saturation density. 
The configurations with trapped antineutrinos and nonvanishing 
$\mu_{\bar\nu_e}$ to compare with we choose to have the same total baryon 
number as the reference star: $N_B = 1.48~ N_{\odot}$, where 
$N_{\odot}$ is the total baryon number of the sun.
For $\mu_{\bar\nu_e}=72$ MeV we obtain $M_A = 1.47~M_{\odot}$ and 
for $\mu_{\bar\nu_e}=150$ MeV, $M_B = 1.72~M_{\odot}$. 
The differences in the radii are  $R_A - R_f= 0.1$ km and $R_B-R_f = 0.3$ km 
and in the central densities $n_q^A-n_q^f = 0.6~n_0$ and 
$n_q^B-n_q^f = 2.1~n_0$, respectively. This is a consequence of the 
hardening of the EoS due to the presence of antineutrinos. 
The mass defect $\Delta M_{if} = M_i - M_f$ can be interpreted as an energy 
release if there is a process which relates the configurations with $M_i$ and
$M_f$ being the initial and final states, respectively. In the right panel of
Fig. \ref{QSConf} we show limits for the binding energy release as a function
of the final state mass when conservative values for the antineutrino 
chemical potential during the trapping are chosen.
The antineutrino untrapping transition results in a first order phase 
transition which gives rise to an explosive release of the binding energy, as
required for a scenario which should explain the engine of supernovae or
gamma-ray bursts \cite{Aguilera:2002dh}.

\subsection{Hybid stars}
Once the EoS with a quark-hadron phase transition is defined and the 
constraints of beta equilibrium and charge neutrality are obeyed (see previous 
Section), the corresponding hybrid star configurations are obtained from the
solution of Einsteins equations \cite{Glendenning:1997wn}.  
\subsubsection{Is a quark core possible?}
\begin{figure}[ht]
\includegraphics[width=.7\textwidth,angle=-90]{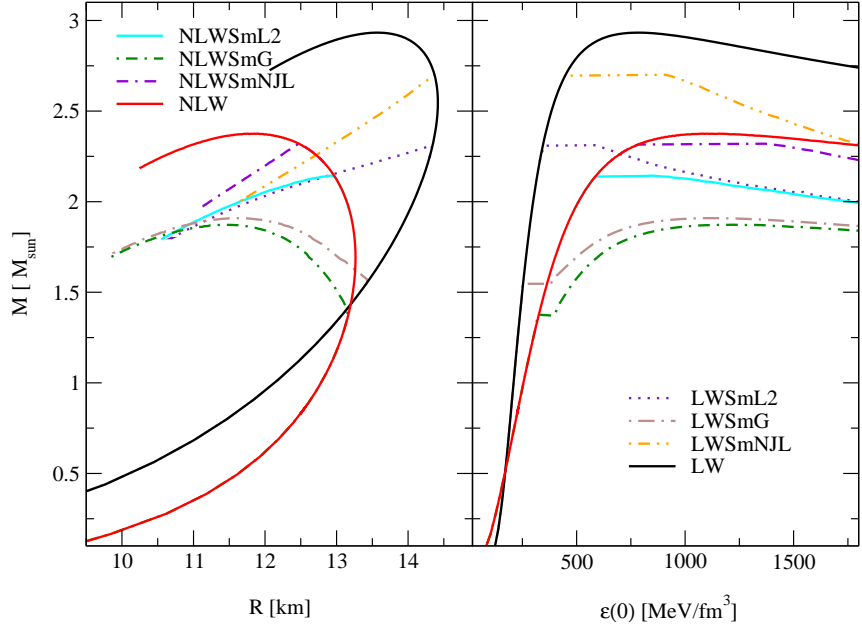}
 \caption{Mass - radius and mass - central energy density curves
for hybrid stars. Stable quark matter cores are possible for a Gaussian 
formfactor model but not for a cut-off one (NJL). }
\label{HSCff}
\end{figure}

The answer to this question depends crucially on the employed EoS. Within the 
setting of the present model, stable quark matter cores are possible for a 
Gaussian formfactor model but not for a cut-off one (NJL). 
This confirms on the one hand conclusions previously obtained within other 
approaches using an NJL model \cite{Schertler:1999xn,Baldo:2002ju} for quark 
matter but on the other hand presents an alternative
quark matter model for which quark cores are possible, see Fig. \ref{HSCff}.

\subsubsection{Cooling curves and the compact star in 3C58}
\begin{figure}[ht]
\includegraphics[width=.7\textwidth,angle=-90]{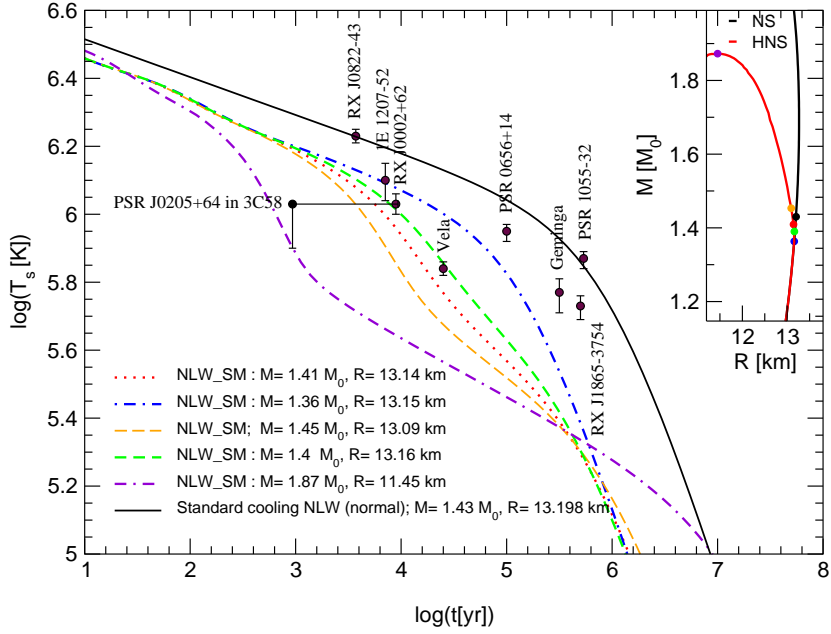}
 \caption{Cooling evolution of hybrid stars with different
masses. The dots in the inset plot of the mass - radius relation correspond to
configurations for which the cooling evolution is shown.  
The observational data are taken from \cite{Yakovlev:2002cs}.}
\label{cool}
\end{figure}

In order to obtain the cooling behavior of the presented hybrid star model,
we employ the program code developed recently 
\cite{Blaschke:1999qx,Blaschke:2000dy} so as to describe 2SC quark matter
and investigate the dependence on the compact star mass \cite{cooling}.
The result shown in Fig. \ref{cool} demonstrates that a massive star
may have a large enough quark core to entail an enhaced cooling behavior in 
accordance with the recent data point reported by CHANDRA measurements
of the pulsar in the supernova remnant 3C58. This is, however, not a unique 
feature of a quark matter interior and could also be explained by other
exotic phases of dense matter \cite{Yakovlev:2002cs}.

\subsubsection{Population gap for accreting LMXBs}
\begin{figure}[ht]
\includegraphics[width=.8\textwidth,angle=0]{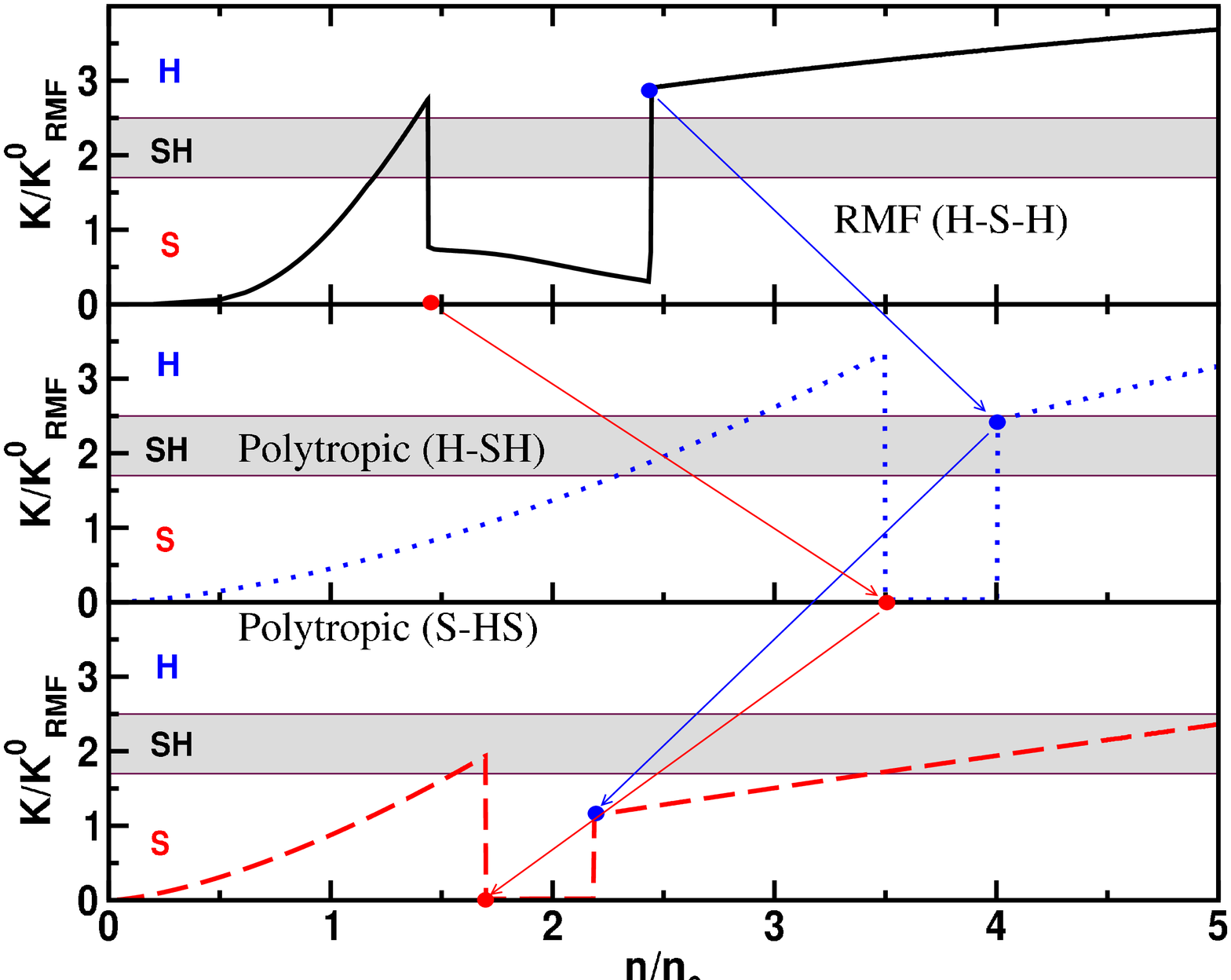}
 \caption{Incompressibilities for Relativistic Mean Field (RMF) and 
polytropic EoS models with a phase transition, see \cite{Blaschke:2002kd}.}
\label{EoSfig}
\end{figure}

Our focus is on the elucidation of qualitative features of 
signals from the high density phase transition in the pulsar timing, 
therefore we use a generic form of an equation of state (EoS) with such a 
transition. 
We use the polytropic type  equation of state  
for different values of the incompressibility \cite{Glendenning:1997wn} 
$K_{L,H}(n)=9~dP/dn$ at the saturation density, see Ref. 
\cite{Blaschke:2002kd}. 
The phase transition between the lower and higher density phases is made 
by the Maxwell construction and compared to a relativistic mean 
field model consisting of a linear Walecka plus dynamical quark 
model EoS  with a Gibbs construction, Fig. \ref{EoSfig}. 

We introduce a classification of rotating compact 
stars in the plane of their angular frequency $\Omega$ and mass 
(baryon number $N$) which we will call {\it phase diagram}. 
In this diagram, configurations with high density matter cores are separated 
from conventional ones by a critical phase transition line. 
The position and the form of these lines are sensitive to changes 
in the equation of state of stellar matter \cite{Blaschke:2001uj}. 

\begin{figure}[ht]
\includegraphics[width=.75\textwidth,angle=0]{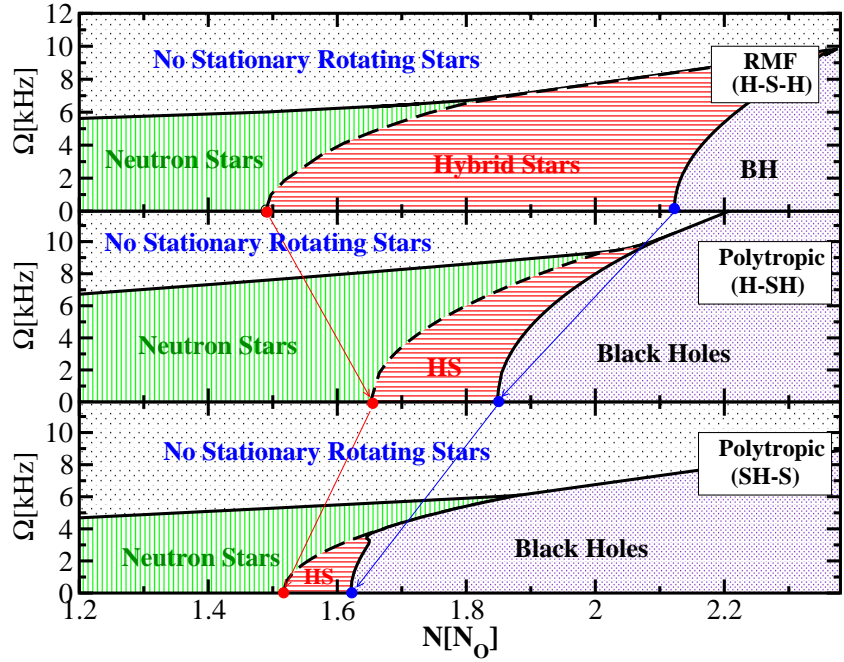}
 \caption{Phase diagrams for rotating star 
configurations corresponding to three model EoS discussed in the text.}
\label{rotate}
\end{figure}

In Fig. \ref{rotate} we display the phase diagrams for the rotating star 
configurations, which correspond to the three model EoS of Fig. \ref{EoSfig}. 
These phase diagrams have four regions: 
(i) the region above the maximum frequency $\Omega >\Omega_{\rm K}(N)$ 
where no stationary rotating configurations are found, (ii) the region of 
black holes $N > N_{\rm max}(\Omega)$, and the region of stable compact 
stars which is subdivided by the critical line $N_{\rm crit}(\Omega)$ into 
(iii) the region of hybrid stars for $N > N_{\rm crit}(\Omega)$ where 
configurations contain a core with a second, high density phase and 
(iv) the region of mono-phase stars without such a core. 

From the comparison of the regional structure of these three different phase 
diagrams in Fig. \ref{rotate} with the corresponding EoS 
in Fig. \ref{EoSfig} we conclude that 
there are the following correlations between the topology 
of the lines $N_{\rm max}(\Omega)$ and $N_{\rm crit}(\Omega)$ and the 
properties of two-phase EoS: 
\begin{itemize} 
\item[-] 
The hardness of the high density EoS determines the maximum mass of 
the star, which is given by the line $N_{\rm max}(\Omega)$. 
Therefore  $N_{\rm max}(0)$ is proportional to the parameter $K_H(n_H)$, 
where $n_H$ is the density of the transition to high density phase. 
\item[-] 
The onset of the phase transition line $N_{\rm crit}(0)$ depends on 
the density $n_H$ and $K_L(n_L)$ where $n_L$ is the density of the transition 
to the low density phase. 
\item[-] 
The curvature of the lines $N_{\rm max}(\Omega)$ and 
$N_{\rm crit}(\Omega)$ is proportional to the compressibility of the high and 
low density phases, respectively. 
\end{itemize} 
Therefore, a verification of the existence of the critical lines 
$N_{\rm crit}(\Omega)$ and $N_{\rm max}(\Omega)$ by observation of 
the rotational behavior of compact objects would 
constrain the parameters of the EoS for neutron star matter. 
We have investigated different trajectories of rotating 
compact star evolution in the phase diagram in order to identify 
scenarios, which result in signatures of the deconfinement phase 
transition \cite{Blaschke:2002kd}.
A key evolutionary track corresponds to accretion with strong magnetic fields 
\cite{Poghosyan:2000mr}. For this case the $\dot \Omega$ first 
decreases as long as the moment of 
inertia monotonously increases with $N$. When passing the critical 
line $N_{\rm crit}(\Omega)$ for the phase transition, the 
moment of inertia starts decreasing and the internal torque term 
$K_{\rm int}$ changes sign. This leads to a narrow dip for $\dot 
\Omega (N)$ in the vicinity of this line. As a result, the phase 
diagram gets overpopulated for $N \stackrel{<}{\sim} N_{\rm 
crit}(\Omega)$ and depopulated for $N \stackrel{>}{\sim} N_{\rm 
crit}(\Omega)$ up to the second maximum of $I(N, \Omega)$ close to 
the black-hole line $N_{\rm max}(\Omega)$. 
A {\it population gap} in the phase diagram of compact stars 
appears as a detectable indicator for hybrid star configurations.

\section{Conclusions}
We have investigated the influence of the diquark condensation on
the thermodynamics of the quark matter under the conditions of
$\beta$-equilibrium and charge neutrality relevant for the discussion of 
compact stars.
The EoS has been derived for a nonlocal chiral quark model
in the mean field approximation, 
and the influence of different formfactors (Gaussian, Lorentzian, NJL) has 
been studied.
We have shown that the smoothness of the interaction changes the 
critical temperatures and chemical potentials for the onset of the phase
transition to lower values. 

The phase transition to color superconducting quark matter from the lower
density regions at small temperatures ($T < 5 \div 10$ MeV) is of first
order, while the melting of the diquark condensate and the corresponding 
transition to normal quark matter at high temperatures is of second order.
The presence of flavor asymmetry due to $\beta$-equilibrium in 
quark matter does not destroy the diquark condensate since the electron 
fraction $n_e/n_{\rm total} < 0.01$ is too small.
The masses of the quark core configurations could be up to
$1.7 ~M_{\odot}$ and the radii could be up to $11$ km.

We have investigated the effects of 
trapped antineutrinos on the asymmetry
and  diquark condensates in a quark star configurations.
By comparing configurations with fixed baryon number the  
release of energy in an antineutrino untrapping transition
is estimated to be of the order of $10^{53}$ erg.
Such a transition is of first order so that antineutrinos can be
released in a sudden process (burst). This scenario could play an important 
role to solve the problem of the engine of 
supernova explosions and gamma ray bursts.
A detailed neutrino transport and cooling calculation 
should be taken into account in a future work. 
A second antineutrino pulse is suggested as an observable characteristics of 
the present scenario.

As there are still many unknowns in the picture we have drawn in this 
contribution for the possible effects of quark matter and color 
superconductivity on compact stars we would like to point out that ther will 
not be one ``smoking gun'' type signal of quark matter but rather the different
facets of the picture which emerges for characteristic features of compact 
stars containing quark matter should all match in the big puzzle. 
To bring the pieces together is a task very similar to that of quark gluon 
plasma search in heavy-ion collision experiments.


\begin{theacknowledgments}
The research of D.N. Aguilera has been supported by DFG Graduiertenkolleg 567
``Stark korrelierte Vielteilchensysteme'', by 
the CONICET (Argentina) and by DAAD grant No. A/01/17862. 
H.G. acknowledges support by DFG under grant No. 436 ARM 17/5/01.
S.Y. was supported by DAAD HOST program during his stay at Rostock 
University. Part of the results reported in these Proceedings have been 
obtained with our collaborators S. Fredriksson, A.M. \"Oztas, G. Poghosyan and 
D.N. Voskresensky. 
\end{theacknowledgments}


\bibliographystyle{aipproc}   

\bibliography{david03}

\IfFileExists{\jobname.bbl}{}
 {\typeout{}
  \typeout{******************************************}
  \typeout{** Please run "bibtex \jobname" to optain}
  \typeout{** the bibliography and then re-run LaTeX}
  \typeout{** twice to fix the references!}
  \typeout{******************************************}
  \typeout{}
 }

\end{document}